\documentstyle[twocolumn,aps]{revtex}
\begin{document}

\title{Negative Resistance and Rectification in Brownian Transport}
\author{Guillermo A. Cecchi and  Marcelo O. Magnasco }
\address{The Rockefeller University, 1230 York Avenue, New York NY10021, USA}
\date{\today}
\maketitle

\begin{abstract}
We discuss under what conditions Brownian transport processes
can display negative resistance.
We prove it cannot occur on 1D spaces like the circle or the line.
We construct an entropic ratchet: an explicit two-dimensional model,
and its collapse onto a branched 1D backbone,
showing negative resistance and rectification as a consequence
of a geometric symmetry breaking.
We derive an accurate numerical method for solving our 2D model.
Finally, we discuss analogies and relevance to
biological ion channels, in particular for channel inactivation and blocking.
\end{abstract}

\narrowtext

If a current arises as a result of some force, then it will flow ``downhill'',
in the direction in which it dissipates energy into heat; thus, the resistance
is always positive.
The diminishing of a current as the driving force becomes stronger is called
negative incremental resistance, or just plain ``negative resistance'' (NR);
devices that display NR exist and have important technological applications;
these devices are, usually, also rectifiers.
If energy is provided through suitable bias voltages, so as to
displace the steady state of operation to the NR region, then the negative
slope of the current can give rise to interesting instabilities like
relaxation oscillations \cite{RELAX,GUK}.
A typical such device from electronics is the tunnel diode.
Much more importantly, the very existance of our nervous systems depends
upon the ability to generate action potentials to transmit nerve impulses
along axons; this phenomenon
requires at least one NR rectifying device, which is known to be
the sodium channel \cite{NR,HH,NA1,NA2,NA3}.

In this Letter, we will show how NR and rectification can arise in a
Brownian transport process,
through a mechanism we will call {\em entropic trapping}. Because 
of the purely geometric nature of this mechanism, we can call our
model an {\em entropic ratchet}. 

We will study a Brownian particle in a periodic potential,
embedded in an equilibrium, constant temperature
bath, and subject to a single force trying to advect it. This type of
process is described by a Langevin equation of the form
$$ {\bf \dot x} = {\bf f(x)} + {\bf \xi}(t) $$
where $\bf x$ belongs to some euclidean space,
$\bf\xi$ is Gaussian white noise satisfying
$ \langle\xi_i(t) \xi_j(s) \rangle = 2 kT \delta_{ij} \delta(t-s)$, and
$\bf f$ is of the form $-\nabla V({\bf x}) + {\bf F}$ with $\bf F$ constant,
i.e., $\bf f$ is a vector field independent of time. We will
assume $V$ to be periodic along the direction of $\bf F$.

Why would it be interesting to search for NR in transport processes of this
form? First of all, these processes have very wide applications
\cite{RISKEN}.
Second, transport in symmetry broken potentials (``ratchets'')
\cite{RISKEN,AP,PRL1,PRL2,MILLONAS,ASTUMIAN,DOERING}
has been shown to be analogous to conduction through electronic diodes,
capable of rectification;
however, the mechanism through which a {\em tunnel} diode provides NR is 
intrinsically quantum mechanical;
hence, it would be interesting to provide classical analog. Finally,
ratchet potentials in one dimension can be shown not to have NR, as we
will now do.

In one dimension, an equation of the form
$$ \dot x = f(x) + \xi(t),\qquad \langle \xi(t)\xi(s)\rangle=2kT\delta(t-s)$$
has an associated stationary Fokker-Planck equation
$$ \partial_t P(x,t) + \partial_x J(x,t)  = 0 $$
$$ J(x,t)= fP(x,t) - kt \partial_x P(x,t) $$

whose steady state, for periodic $f(x)$, can always be solved in quadratures.
If $f$ has a zero spatial
average, then it is the gradient of a periodic potential, and detailed
balance and Boltzmann weights hold. If $f$ does not have a zero average,
then it can be written as $f = -\partial_x V + F $ with $V$ periodic,
and the stationary state can be solved in double quadratures \cite{PRL2}.
The Fokker-Planck probability current $J$ as a function of $F$ is given by:
\begin{eqnarray}
J(F) & = & {kT (e_{kT}^{2\pi F} -1) \over Q(F)} \\
Q(F) & = & \oint \oint e_{kT}^{ V(x')-V(x) + F (x-x') +2\pi F \Theta(x'-x) } dx' dx
\end{eqnarray}
where $e^x_{kT}$ means $\exp(x/kT)$.
$Q$ is positive; moreover, $\partial_F Q$
is also positive, and satisfies $kT \partial_F Q < 2\pi Q $.
This inequality implies $dJ/dF > 0$ and hence Brownian transport in
one-dimensional periodic potentials (under a steady force and in white noise)
cannot show NR. (We will show later that this only holds if the underlying
space is topologically trivial).

Now we will construct an explicit example in two dimensions.
First we illustrate the notion of entropic barriers, through a very simple
example. Let us consider the following potential
$$ V(x,y) = x^2 \left ( \cos(y) + 1.1 \right) $$
We see that the potential is identically zero on the $y$ axis, and
bigger than zero everywhere else. Hence there are no true energy barriers
impeding motion of a Brownian particle along the periodic direction $y$.
However, the shape of the potential around the $y$ axis is also important
at nonzero temperature. For any given $y$, a slice of the potential along
that value of $y$ is a parabola; however, as motion progresses along $y$ this
parabola opens and closes periodically. In the absence of an external force,
a Brownian particle will spend more time around $y = \pi$ than on the
bottleneck $y=0$, every now and then jumping one period up or down in $y$, as
if there actually were an energy barrier; these are called {\it entropic
barriers} because, unlike a true activation energy, the timescales they
induce do not follow Arrhenius-Kramers laws. In other words, as will see,
the violation of these laws is responsible for the non-monotonic
behavior of the current as a function of the applied force $F$.

We will now break the parity symmetry along the $y$ axis; if we loosely call
any parity broken potential a ``ratchet'', we can say the following
potential is an {\em entropic ratchet:}
\begin{equation}
 V(x,y) =  x^2 \left ( \cos ( y + \ln\cosh x ) + 1.1 \right) /2
\label{pot}
\end{equation}
where $\ln\cosh x$ is just an easy way to make a function that is both even
and linear in $x$ for large $x$.
The equipotentials are now symmetry broken and look like a
herringbone pattern, see Fig. 1.
Our dynamics will be given by
\begin{equation}
 {\bf \dot x} = -\nabla V + F {\bf\hat y} + {\bf \xi}(t)
\label{lang}
\end{equation}
where $\bf\hat y$ is the unit vector along $y$, and the noise correlators
are as before.

So, if we apply a force $F$ along $y$, if the force is positive the particle
will move forward without problem. However, if the force is negative, the
particle will move backwards, but every now and then it will get into a
spine of the herringbone. It will progress upon the spine for a while,
deeper the stronger the force, and then will have to go {\em against} the
force in order to climb back up and get again on the backbone. Thus it will
be locked for a while, because the energy required to climb back up is a
honest activation energy, and the time required to do so obeys an Arrhenius
law. But the probability that a particle on the backbone will get into the
spine does not depend exponentially on $F$, and so
the particle spends a larger proportion of time blocked away in the spines
as $F$ becomes more negative. The overall timescale to get into
and out the spine is not Arrhenius-like, yet near the bottleneck
the potential energy of points
in the spine and in the backbone is the same. If it was, the time the
particle spends stuck in the spine would cancel out with the time it spends
in the backbone, and the current would be a monotonic function of $F$.


We have evolved numerically Equation \ref{lang} to obtain the average speed
of the particle as a function of $F$ and $kT$. The mean speed equals the
Fokker-Planck current times the period ($2\pi$) of the potential. The
result of our simulations is shown in Figure 2.

In performing a numerical integration of Equation \ref{lang}, one encounters
several numerical problems. The simplest method for numerical integration
is the Euler method,
$$ x(t+\Delta t) = x(t) + f(x) \Delta t + \sqrt{2 kT \Delta t} \eta $$
where the $\eta$ are random gaussian numbers with unit variance. While the
Euler method is first order in $\Delta t$ for an ordinary differential
equation, it is only one-half-order for a Langevin equation, a property
characteristic of diffusion. This implies that extremely small timesteps
have to be used for accurate integration.
Several methods have been proposed
for increasing the order of the integration, including stochastic versions
of Runge-Kutta \cite{SRK}.
However, the spines in our potential are a {\em stiff}
problem, being long and skinny, which would limit the applicability of a
Runge-Kutta scheme even in the absence of noise. The reason is that the
timescale has to be small in comparison with the relaxation time on the
fastest direction, in this case the skinny direction normal to the axis
of the spine. But then the interesting time evolution is that {\em along}
the axis of the spines, which becomes painfully slow.

We have devised a method to cope with this problem. The fundamental problem in
developing a Runge-Kutta scheme is that somehow one is assuming analyticity
both of the vector field and of the solution; the last one is just not there.
But one can safely assume analyticity of the vector field alone.
If we expand the latter to first order,
$$\bf f(x+\Delta x)\approx f(x) + \Delta x \cdot (\nabla f)(x)$$
We use the fact that $\bf f$ is curl-free (and thus $\bf\nabla f$ is
a symmetric tensor) to diagonalize it and rotate to its eigenbasis.
In this base, the problem locally becomes
a cross-product of independent Ornstein-Uhlenbeck processes. Thus, the
question is: if we know we currently are at position $x$ at time $t$,
what is the probability that we will be at position $x'$ at time $t'$?
The answer to this question is the Fokker-Planck propagator (Green function),
which is known analytically for the Ornstein-Uhlenbeck process \cite{RISKEN}.
It is a Gaussian, centered at the position $x'$ that would be the solution for
the deterministic case at time $t'$, and with a width which has been changed because of
the compression/expansion due to $f'\equiv df/dx$.
Thus it is a trivial matter to
generate a new value $x'$ with the correct probability distribution for
an Ornstein-Uhlenbeck process; our numerical method then reads
$$ x' = (x - x_c) e^{-f' \Delta t} + x_c +
\left( (kT/f') (1 - e^{-2 f' \Delta t} ) \right)^{1/2} \eta $$
$$ x_c = x - f/f', \qquad \Delta t = t' - t $$
for each eigendirection of the Hessian. The new coordinates are then
rotated back to the original frame.
The advantages of this method
over stochastic Runge-Kutta are two. First, it solves {\em exactly}
the Ornstein-Uhlenbeck process, by construction, and hence any linear problem.
Second, it can handle stiff problems more easily;
it will not lose accuracy if the timestep is larger than the relaxation
timescale of the fast direction,
because it will not {\em overshoot} and generate dynamical instabilities; one
can concentrate on the more interesting slow timescale.
The disadvantages are also two: analytic knowledge of the Hessian is
required, and a matrix eigensystem calculation has to be performed, together
with two coordinate frame transformations. Thus our method rapidly loses
ground to stochastic Runge-Kutta for high dimensionalities, unless the problem
is quite stiff or the Hessian is sparse.
For our stiff problem in two dimensions, this method is
extremely well adapted; Fig 2 was generated in a few hours of cpu of a
workstation and shows almost no trace of the residual noise typical of
direct Langevin simulations.

Back to our problem, we would like to get some more explicit understanding
of the $J(F)$ curve we have just obtained numerically. Let us consider a
one-dimensional cartoon of the system: we can represent the backbone as
a circle, with a single spine of length $L$ attached to it,
at an angle $\theta$, as in Fig. 3a.
There is no potential,
just the steady force. At the vertex, we have to impose that
the probability be continuous: the limit of $P(x)$
as we approach the vertex from all {\em three} sides must be the same,
$P_{vert}$. The current should be conserved at the vertex, so that
probability does not accumulate. Then, the probability density
on the circle is constant, and so is the current on the backbone:
$J=F P_{vert}$.
The current on the spine should vanish, and hence the probability
along the spine $P(x)\approx \exp(-F x /\tan(\theta))$. Therefore, the final
current is
\begin{equation}\label{cart}
{ F \over J(F)} = {  2\pi +
{kT \over F \cos\theta  }
( 1 - e_{kT}^{-F L \cos\theta }) }
\end{equation}
which has the right qualitative form (Fig 3b),
except for a single detail.
For $F$ large and positive both cartoon and full case
converge algebraically to unity. But for $F$
large and negative, $J\to 0$ exponentially, while in the full case it seems
to die faster. This behavior can be understood if we recall that the spines
are truly different: first, there is a
potential along them; second, they are not of
length $L$, but rather arbitrarily deep. The center of the spines lies
approximately at $y = \pi- \ln\cosh x$, and the
potential there is $V=0.05*x^2$.
The current in this case equals
\begin{equation}
{F \over J(F)} = {  2\pi + 2
\int_{0}^{\infty} \sqrt{1 + \tanh^2(x)} e^{-0.05 x^2 + F \ln\cos(x)} dx}
\end{equation}
Thus, for a large and negative force, the particle entering the spine will
find a stable fixed point at $x \approx 10 F $. Thus the effective length
of the spine {\em increases} as $F$ becomes more negative;
the activation barrier is the product of this effective length and $F$, and
hence {\em quadratic} in $F$ rather than linear. So we have shown that NR
is actually possible in one dimension, but only if the topology of the
space is more complicated than just a circle.


We have presented an explicit example of a Brownian transport
process showing NR, The NR region is generated through an essentially
entropic process. There are no obstacles, no energy barriers to the motion
of the particle along the $y$ axis. There is just a finite probability of
exploring space a bit out of the center, and end up blocked inside a spine.
Since we are in a thermal, rather than quantum, situation, our system has
much more similarity to biological ion channels than to electronics.
There are two well known instances in channels where NR is observed.
The first one is in channels that can have an {\em inactive} state; this
is the case of the Na channel, but some K channels also have this property;
this is extremely important biologically, because the NR resulting from
inactivation is essential to the regeneration of action potentials.
The second one is the blocking of channels through large ions; this is
important experimentally, because in order to assess properties of new
channels, biophysicists will test for changes in behavior when the channel
is ``poisoned'' with various compounds of known effect on known channels.
The standard poison arsenal includes several large ions, that can get partially
into the channel and block it; for example,
tetraethylammonium is used for Na channels \cite{TEA} and Mg \cite{K},
Cs \cite{CS} or polyamines \cite{K2} for K channels.
It is worth noting that the I-V curves of such channels look extremely similar
to that of our model, and, furthermore,
several of the experimental measurements \cite{NA2,TEA,CS}
show the faster-than-exponential decay of our model, while standard
theoretical models with {\em fixed} barriers \cite{K,BROLI}
show exponential decays;
this discrepancy can only be solved through models having barriers that 
depend on the field.
This might mean that the large ion buries itself deeper and deeper into
the crevice of the channel, getting more and more stuck and having to climb
a larger distance against the potential to get out, just as for our spines.
There are also similarities with the case of channel inactivation.
These channels are hypothesized to have three states: closed, open,
and inactive. The transition between open and inactive states
has been modeled with a ``ball and thread'' mechanism \cite{NA3,BROLI,BAT}:
some mobile, charged, globular piece of the channel loosely attached
through a long thread can get stuck in the mouth of the channel and block
it; pretty much like a large ion blocker.  Our model does not
include a closed state, since the central backbone is always ``open'';
thus we do not see the current going to zero as $F$ becomes large and
positive.
The closed state can be bypassed by looking at the peak sodium current from
a pulse rather than the stationary current \cite{NA4}, and in this case
the observed peak I-V curve is qualitatively similar to what we observe.

We wish to thank Albert Libchaber, Sanderman Simon and Gustavo Stolovitzky
for many discussions.


\begin{figure}
\caption{The equipotentials of the model.}
\label{equi}
\end{figure}
\begin{figure}
\label{res}
\caption{ $J(F)$ at $kT=1$ as computed from Langevin simulations.
Each run lasted for $10^5$ units of time; ten runs were done for each
value of $F$. }
\end{figure}
\begin{figure}
\label{cartoon}
\caption{(a) The phase space for the cartoon. (b) $J(F)$ as given by
Eq. 4}
\end{figure}


\begin{thebibliography}{99}
\bibitem{RELAX}  N. Minorsky, {\sl Nonlinear Oscillations}, Van Nostrand,
Prin-ceton (1962).
\bibitem{GUK} J. Guckenheimer and P. Holmes,
{\sl Nonlinear Oscillations, Dynamical Systems,
and Bifurcations of Vector Fields}, Springer-Verlag, Berlin (1983).
\bibitem{NR} Kornacker, K., in {\sl Biological Membranes}, Dowben, R. M, ed.,
Little, Brown and Co., Boston (1969)
\bibitem{HH} Hodgkin, A. L. and Huxley, A. F. {\sl J. Physiol. (Lond.)}
{\bf 116}: 449-472, 473-496; 497-506; {\bf 117}: 500-544 (1952)
\bibitem{NA1} Hille, B., {J. Gen. Physiol.} {\bf 66} 535-560 (1975)
\bibitem{NA2} Dubois, J. M., Schneider, M. F. and Khodorov, B. I.,
{\sl J. Gen. Physiol.} {\bf 81} 829-844 (1983)
\bibitem{NA3} B. Alberts, D. Bray, J. Lewis, M. Raff, K. Roberts and
J. D. Watson, {\em The Molecular Biology of the Cell}, ($2^{nd}$ edition),
Garland, New York (1989)
\bibitem{RISKEN} H. Risken, {\sl The Fokker-Planck Equation: Methods of Solution
and Applications}, Berlin: Springer-Verlag (1989)
\bibitem{AP} A. Ajdari and J. Prost,
{\sl C. R. Acad. Sci. Paris II}, {\bf 315} 1635 (1993)
\bibitem{PRL1} M. Magnasco, {\sl Phys. Rev. Lett.} {\bf 71}.10, 1477 (1993)
\bibitem{PRL2} M. Magnasco, {\sl Phys. Rev. Lett.} {\bf 72}.16, 2656 (1994)
\bibitem{MILLONAS} M. Millonas and M. Dykman, {\sl Phys. Lett. } {\bf A185} 65
(1994)
\bibitem{ASTUMIAN} D. Astumian and M. Bier,
{\sl Phys. Rev. Lett.} {\bf 72} 1766 (1994)
\bibitem{DOERING} C. Doering, W. Horsthemke and J. Riordan, {\sl Phys. Rev.
Lett. } {\bf 72}, 19, 2984 (1994)
\bibitem{SRK} Honeycutt, R. L. {\sl Phys. Rev. A} {\bf 45}.2 600-603 (1992)
\bibitem{TEA} Armstrong, C. M. and Binstok, L. {\sl J. Gen. Physiol.}
{\bf 48} 855-872 (1965)
\bibitem{K} Hille, B. and Schwartz, W.,  {\sl J. Gen. Physiol.}
{\bf 72} 409-442 (1978)
\bibitem{CS} Hagiwara, S., Miyazaki, S. and Rosenthal, N. P.,
{\sl J. Gen. Physiol.} {\bf 67} 621-638 (1976)
\bibitem{K2} Lopatin, A. N., Makhina, E. N. and Nichols, C. G.,
{\sl Nature} {\bf 327} 366-369 (1994)
\bibitem{BROLI} Hille, B., {\sl Ionic Channels of Excitable Membranes},
Sunderland MA: Sinauer Associates (1984)
\bibitem{BAT} Armstrong, C. M. and Bezanilla, F. {\sl J. Gen. Physiol.}
{\bf 70} 577-590 (1977)
\bibitem{NA4} Hille, B., Woodhull, A. M. and Shapiro, B. I.
{\sl Phil. Trans. R. Soc. Lond. B} {\bf 270} 301-318 (1975)



\end{thebibliography}
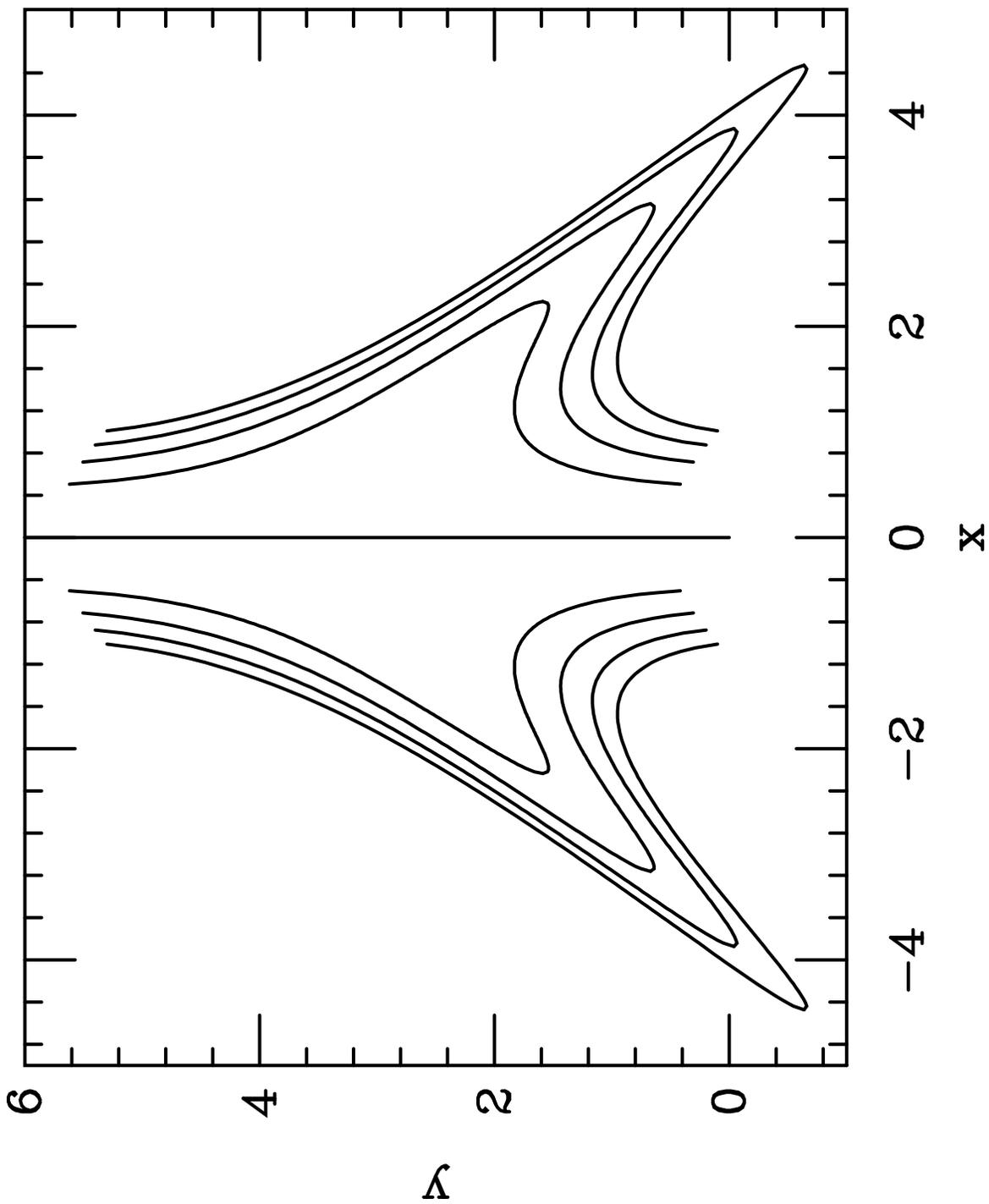
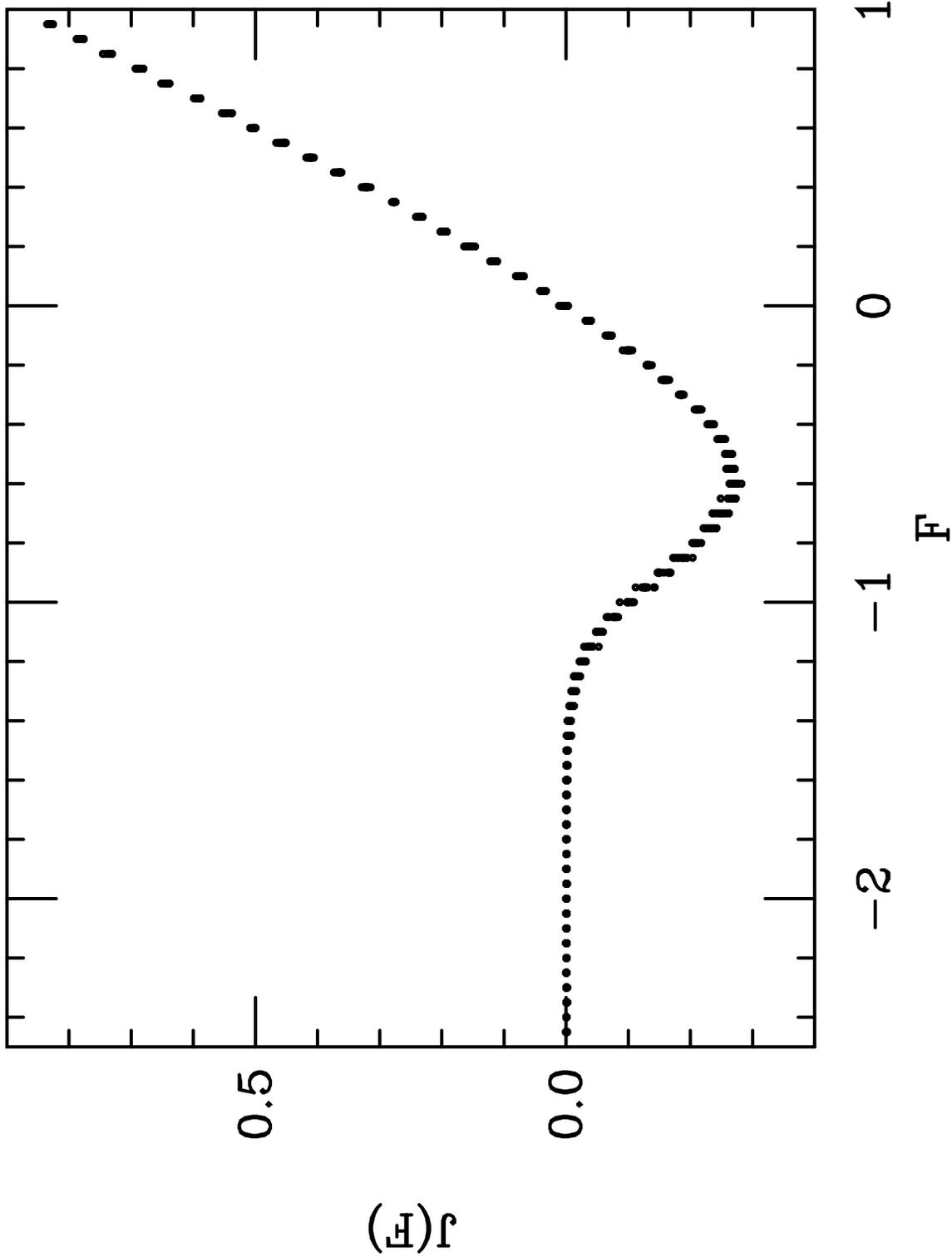
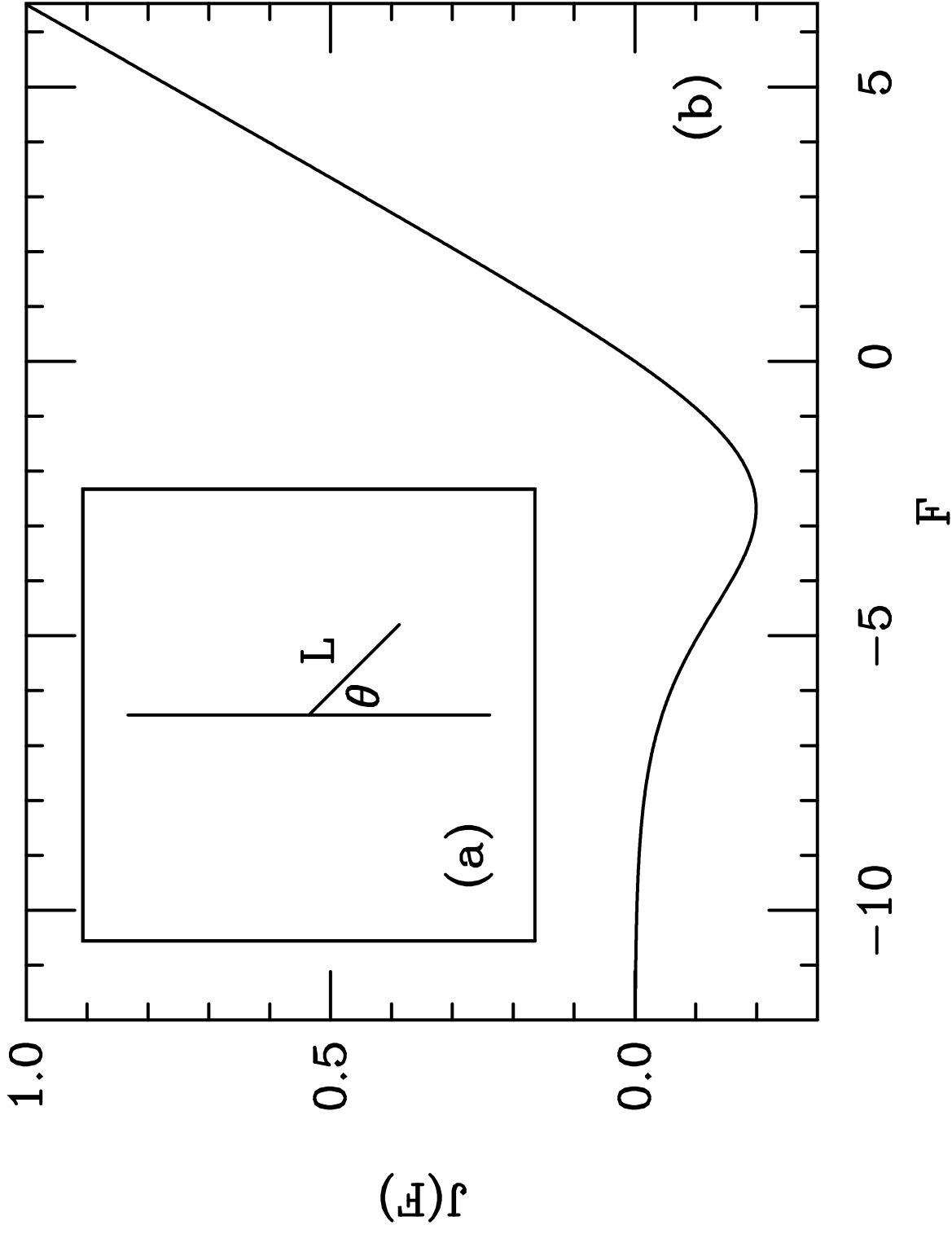
\end{document}